\documentclass[runningheads]{llncs}

\usepackage[T1]{fontenc}
\usepackage{graphicx}
\usepackage{bbding}
\usepackage[colorlinks=true,allcolors=black]{hyperref}
\usepackage{xurl}
\usepackage{hyperxmp}
\hypersetup{
  pdftitle={HERITRACE: a domain-agnostic framework for SHACL-driven RDF curation with provenance and change tracking},
  pdfauthor={Arcangelo Massari, Silvio Peroni},
  pdfsubject={Open-source domain-agnostic web application for SHACL-driven RDF data curation with provenance documentation and change tracking},
  pdfkeywords={Semantic Web, RDF, data curation, provenance, change tracking, SHACL, cultural heritage, usability evaluation, open source, FAIR},
  pdfcopyright={Creative Commons Attribution 4.0 International},
  pdflicenseurl={https://creativecommons.org/licenses/by/4.0/},
  pdflang={en},
}
\usepackage{listings}
\usepackage{tabularx}
\usepackage{amssymb}
\usepackage{xcolor}

\lstdefinelanguage{turtle}{
  morekeywords={@prefix, a},
  sensitive=true,
  morecomment=[l]{\#},
  morestring=[b]",
}

\begin{document}

\title{HERITRACE: a domain-agnostic framework for SHACL-driven RDF curation with provenance and change tracking}

\titlerunning{HERITRACE: a domain-agnostic framework for RDF curation}

\author{Arcangelo Massari\inst{1}\orcidID{0000-0002-8420-0696}\Envelope \and
\\Silvio Peroni\inst{1,2}\orcidID{0000-0003-0530-4305}}

\authorrunning{A. Massari, S. Peroni}

\institute{Research Centre for Open Scholarly Metadata, Department of Classical Philology and Italian Studies, University of Bologna, Bologna, Italy\and Digital Humanities Advanced Research Centre (/DH.arc), Department of Classical Philology and Italian Studies, University of Bologna, Bologna, Italy\\
\email{\{arcangelo.massari,silvio.peroni\}@unibo.it}}

\maketitle

\begin{abstract}
HERITRACE is an open-source web application that enables users without Semantic Web expertise to curate RDF data through form-based interfaces with automatic provenance documentation and change tracking in RDF. It uses SHACL for data model definition and form generation, connects to existing SPARQL-accessible stores without data migration, and records every modification as a provenance snapshot that can be browsed and restored. HERITRACE is domain-agnostic and its configuration is layered: a minimal deployment only connects to a SPARQL store, while SHACL shapes and YAML display rules refine the data model and presentation, and IRIs are minted automatically by default or through a Python class for institution-specific schemes. This paper describes the software architecture and provides the first empirical evaluation. HERITRACE is deployed in production for the ParaText project, where classical philologists curate bibliographic data about ancient Greek exegetical traditions, and is planned as the editing interface for OpenCitations and as the curation layer for the Social Sciences and Humanities Citation Index within the GRAPHIA Horizon Europe project. Since it operates on any SPARQL-accessible store without data migration, its adoption potential extends to any domain maintaining RDF data. HERITRACE is publicly available on GitHub under the ISC license, archived on Zenodo and Software Heritage Archive, and documented for deployment with a pre-built Docker image.

\end{abstract}

\section{Introduction}\label{sec:introduction}

Cultural heritage institutions increasingly use Semantic Web technologies to publish their cultural heritage collections on the Web~\cite{tamperTextKnowledgeMethods2023}. RDF~\cite{manolaRDFPrimer2004} and Linked Open Data (LOD) principles~\cite{bizerLinkedDataStory2009} allow them to share metadata in machine-readable formats that are Findable, Accessible, Interoperable, and Reusable (FAIR)~\cite{wilkinsonFAIRGuidingPrinciples2016}. However, this adoption has created a tension between data quality and technical expertise. Data quality depends on domain knowledge for relationship validation, metadata enrichment, and entity disambiguation, yet the systems managing this data require proficiency in SPARQL~\cite{w3cSPARQL11Query2013} and OWL~\cite{w3cOWL2Web2012} that domain experts typically lack. A second gap concerns how systems represent the history of the data: tools that do provide user-friendly interfaces typically record provenance and change tracking outside RDF, in relational stores, so this metadata does not receive the same FAIR treatment as the data it documents.

Two existing cases can be briefly introduced to illustrate these gaps. The Digital Library of the Department of Classical Philology and Italian Studies (FICLIT) at the University of Bologna~\cite{adlab-laboratorioanalogicodigitaleFICLITDigitalLibrary2022} employs OmekaS~\cite{rueffDealingPostexcavationData2024} to offer user-friendly interfaces but records provenance outside RDF, lacks change tracking, and requires CSV-based bulk loading for RDF import, preventing FAIR management of data and metadata. OpenCitations~\cite{peroniOpenCitationsInfrastructureOrganization2020} (\url{https://opencitations.net}), an open scholarly infrastructure publishing open bibliographic metadata and citation data, implements a complete RDF representation for both its data and provenance, but requires technical expertise for data quality management, excluding librarians who could identify errors in bibliographic metadata. A solution that maintains complete provenance and change tracking within RDF while remaining usable by domain experts would be necessary for properly addressing both cases presented.

This paper presents HERITRACE, an open-source web application~\cite{arcangelomassariOpencitationsHeritraceV3002026,massariHERITRACESoftwareHeritage2026} formally maintained by OpenCitations, which ensures its sustainability by having HERITRACE as part of its tools and services. HERITRACE enables domain experts to curate RDF data through form-based interfaces while automatically maintaining provenance documentation and change tracking, with direct integration into existing RDF collections. A previous publication~\cite{massariHERITRACEUserFriendlySemantic2025} introduced the system's design rationale. Since then, the system has been extended with additional functionalities: virtual properties that translate complex RDF patterns into fields that match users' mental models, configurable orphaned-entity handling, pluggable IRI minting, shape-level configuration binding that allows distinct interfaces for entities sharing the same RDF class, and context-sensitive validation through SHACL conditional constraints. Beyond these extensions, this paper contributes a description of the software architecture, a refined functional comparison with related systems, and the first empirical evaluation of HERITRACE with real users.

The rest of the paper is organised as follows. Section~\ref{sec:relatedworks} reviews existing systems for semantic data management. Section~\ref{sec:architecture} describes the HERITRACE framework in detail. Section~\ref{sec:evaluation} assesses HERITRACE through functional comparison, a usability study, and community adoption. Section~\ref{sec:limitations} discusses the findings alongside limitations and future work.

\section{Related Works}\label{sec:relatedworks}

Giacomini \textit{et al.}~\cite{giacominiCLEF20Solutions2025} surveyed existing systems for the management of linked open data in cultural heritage. This section builds on their survey.

\textbf{OmekaS}~\cite{rueffDealingPostexcavationData2024} provides a user-friendly web-publishing platform primarily serving museums and educational institutions. It stores both data and provenance metadata in a relational database, making it architecturally incompatible with operating directly on an existing quadstore. Moreover, it does not maintain change histories, and importing RDF data requires bulk loading through CSV conversion~\cite{berthereauCSVImport2024}.

\textbf{Semantic MediaWiki}~\cite{krotzschSemanticMediaWiki2006} integrates semantic capabilities into the MediaWiki platform, inheriting its page-level revision history for provenance and change tracking, though this information is stored in MediaWiki's relational database rather than in RDF. RDF import through the RDFIO extension~\cite{lampaRDFIOExtendingSemantic2017} supports only triples, not named graphs.

\textbf{ResearchSpace}~\cite{oldmanReshapingKnowledgeGraph2018} provides diverse visualization options and records when modifications occur and which agents made them. However, it does not record what specific changes were made. Configuring the interface requires creating templates in HTML, Handlebars, and ResearchSpace-specific components rather than standard or widely adopted languages, imposing a steep learning curve on administrators. It supports direct RDF upload but requires significant programming to adapt the interface after import.

\textbf{CLEF}~\cite{daquinoCLEFLinkedOpen2023,giacominiCLEF20Solutions2025} manages digital libraries with provenance via named graphs and change tracking through GitHub synchronization, but lacks a graphical restoration interface. It supports only individual item creation, not bulk RDF import: integrating a pre-existing triplestore requires restructuring all data into CLEF's per-record named graphs, each combining entity data with provenance metadata.

\textbf{Wikibase}~\cite{diefenbachWikibaseInfrastructureKnowledge2021}, the software powering Wikidata, combines user-friendliness with revision history inherited from MediaWiki, though revision data is stored in MediaWiki's relational database rather than represented in RDF. It restricts predicates to its internal Property namespace, preventing direct reuse of external OWL ontology terms and requiring conversion through tools such as O2WB~\cite{dobriyO2WBToolEnabling2023}.

\textbf{Sinopia}~\cite{schreurSinopiaLinkedDataEditing2019} generates form-based interfaces from JSON profile templates and, although primarily used for BIBFRAME workflows, supports properties from any RDF ontology. Data is stored as JSON-LD in MongoDB rather than in a triplestore, and administrative metadata recording who modified a resource and when is not represented in RDF. Previous versions of resources can be viewed but not restored. RDF data can be imported by pasting serialized RDF into a text area, but the system cannot connect to external SPARQL-accessible stores, limiting scalability for large collections.

\textbf{Schímatos}~\cite{10.1007/978-3-030-62466-8_5}, like HERITRACE, builds web forms from SHACL shapes. It also lets developers create and edit shapes through a graphical interface, a capability HERITRACE delegates to file-based configuration. However, Schímatos manages neither provenance nor change tracking. Moreover, it provides no presentation layer to abstract the data model: entity references appear as IRIs unless \texttt{rdfs:label} triples are added to the data, since display names cannot be configured from the graph, and reified relationships are surfaced as stored rather than abstracted into the fields a curator expects.

\section{The HERITRACE Framework}\label{sec:architecture}

The design of HERITRACE is guided by five requirements derived from the comparison criteria introduced by Giacomini \textit{et al.}~\cite{giacominiCLEF20Solutions2025} and a survey of OpenCitations personnel needs for making its data modifiable by domain experts, such as librarians, while retaining the existing backend technologies.

From Giacomini \textit{et al.}, three criteria are adopted: (R1)~user-friendliness for end users, (R2)~user-friendliness for administrators, and (R3)~provenance management. Two criteria are refined: ``data management integration'', which mentions versioning without defining specific capabilities, becomes (R4)~change tracking, requiring that modifications are recorded and versions can be restored; ``reusability'', which was too broad, becomes (R5)~direct RDF import, requiring connection to pre-existing RDF collections without data migration. R3 and R4 are evaluated specifically as RDF-based capabilities: provenance and change tracking metadata should receive the same FAIR-compliant treatment as the data they document, rather than being stored in relational databases outside the Linked Open Data ecosystem. 

These requirements are further justified by OpenCitations' needs~\cite{peroniOpenCitationsInfrastructureOrganization2020}. Its collections are natively stored in RDF: the OpenCitations Meta collection~\cite{opencitationsOpenCitationsMetaCSV2026} contains over 129~million bibliographic entities, and the OpenCitations Index collection~\cite{opencitationsOpenCitationsIndexCSV2026} over 2.4~billion citations. At this scale, restructuring the data is impractical, making direct RDF import (R5) a prerequisite. Provenance management (R3) and change tracking (R4) are particularly relevant because OpenCitations aggregates data (as of 2 May 2026) from five sources -- Crossref~\cite{hendricksCrossrefSustainableSource2020}, DataCite~\cite{braseTenthAnniversaryAssigning2015}, the NIH Open Citation Collection~\cite{hutchinsNIHOpenCitation2019}, the Japan Link Center~\cite{jalcJapanLinkCenter}, and OpenAIRE~\cite{manghiOpenAIREplusEuropeanScholarly2012} -- and another source, i.e. OpenAlex~\cite{priemOpenAlexFullyopen2022}, is used for alignment purposes~\cite{rizzettoMappingBibliographicMetadata2024}. Tracking which source contributed each datum, when, and recording any subsequent corrections is necessary for data verifiability and reliability.

Regarding the development methodology, test-driven development~\cite{beckTestdrivenDevelopmentExample2003} guided the implementation. The test suite produced during development also includes integration tests that exercise the system against real triplestore instances, testing state combinations that unit tests cannot reach. A continuous integration pipeline~\cite{humbleContinuousDeliveryReliable2011} executes the full test suite on every commit via GitHub Actions across various Python versions. Releases follow semantic versioning~\cite{prestonwernerSemanticVersioning2013} and change logs are generated from Conventional Commits~\cite{conventionalCommits2019}. Machine-readable software metadata is provided through a CodeMeta~\cite{codemetaProject} descriptor, and per-file licensing information is made machine-readable through the REUSE Specification~\cite{fsfeREUSESpecification2024}; compliance is enforced in continuous integration.

\subsection{Software Architecture}\label{subsec:overview}

\begin{figure}[!ht]
\centering
\includegraphics[width=\textwidth]{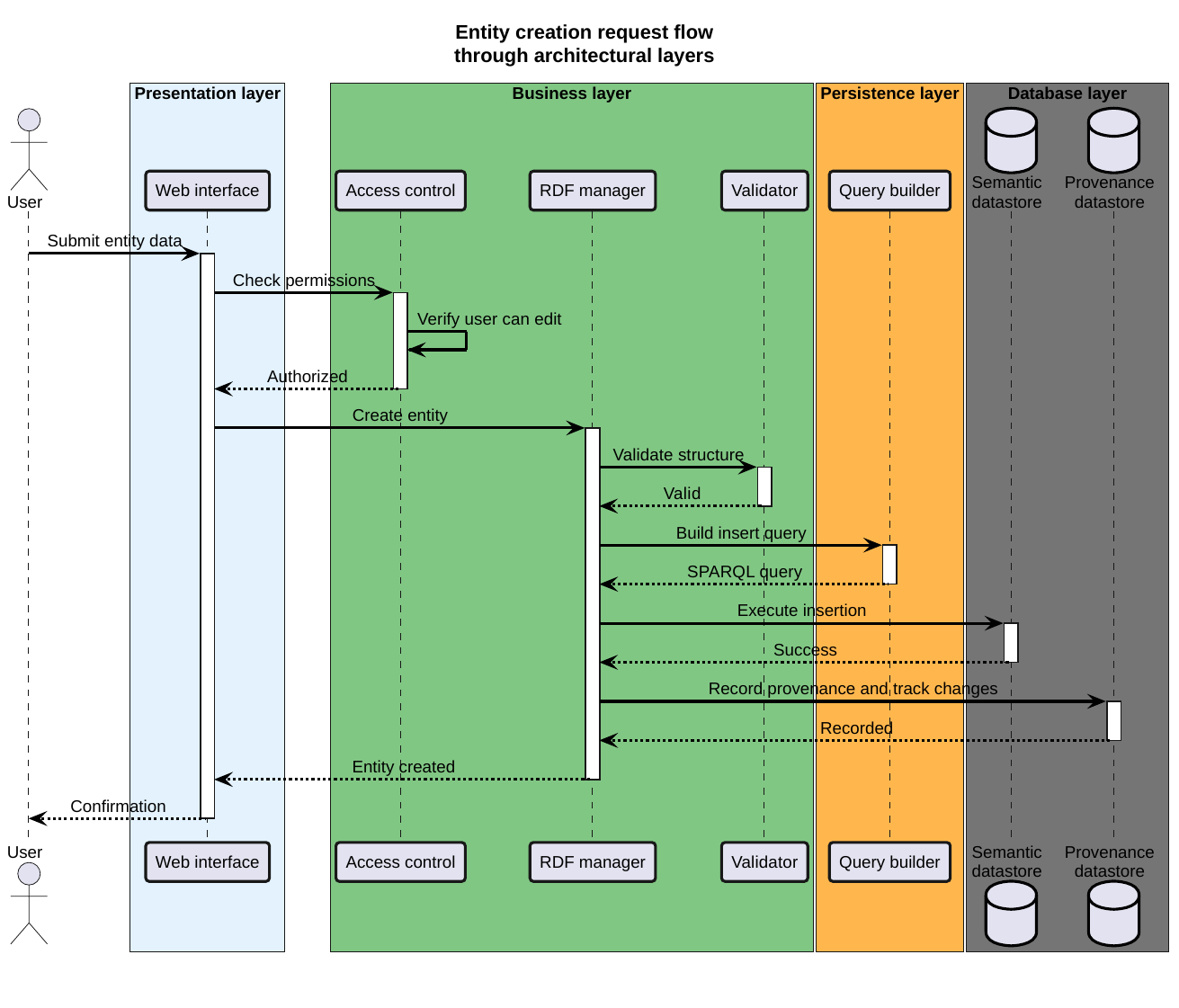}
\caption{Entity creation request flow through HERITRACE architectural layers}\label{fig:request_flow}
\end{figure}

HERITRACE implements a layered architecture~\cite{richardsFundamentalsSoftwareArchitecture2020} with presentation, business, persistence, and database layers (Fig.~\ref{fig:request_flow}), processing each request through these layers sequentially. In the business layer, Access control verifies editing permissions, the RDF manager coordinates entity operations, and the Validator checks data against SHACL~\cite{knublauchShapesConstraintLanguage2017} shapes. In the persistence layer, the Query builder translates these operations into SPARQL queries for the underlying stores.

The database layer contains two SPARQL-accessible stores: the Semantic datastore holds entity data (supporting both triplestores and quadstores) and the Provenance datastore holds versioning data (requiring a quadstore for named graph-based snapshots). The two stores can be configured as separate databases or a single shared instance. A Redis~\cite{sanfilippoRedis2009} state store manages concurrent edit locks. Users authenticate through ORCID~\cite{haakORCIDSystemUniquely2012} OAuth~2.0, with an allowlist of ORCID identifiers restricting editing access to authorized personnel. 

This architecture keeps each layer independent: the database layer operates with any SPARQL-accessible store, from Virtuoso~\cite{erlingRDFSupportVirtuoso2009} to QLever~\cite{bastQLeverQueryEngine2017}, and the presentation layer could be replaced with a framework for building multi-platform applications like Flutter~\cite{googleFlutter2026} without modifying the business logic.

\subsection{Domain-Agnostic Data Editing}\label{subsec:editing}

\begin{figure}[!ht]
\centering
\begin{minipage}[t]{0.48\textwidth}
\centering
\includegraphics[width=\textwidth]{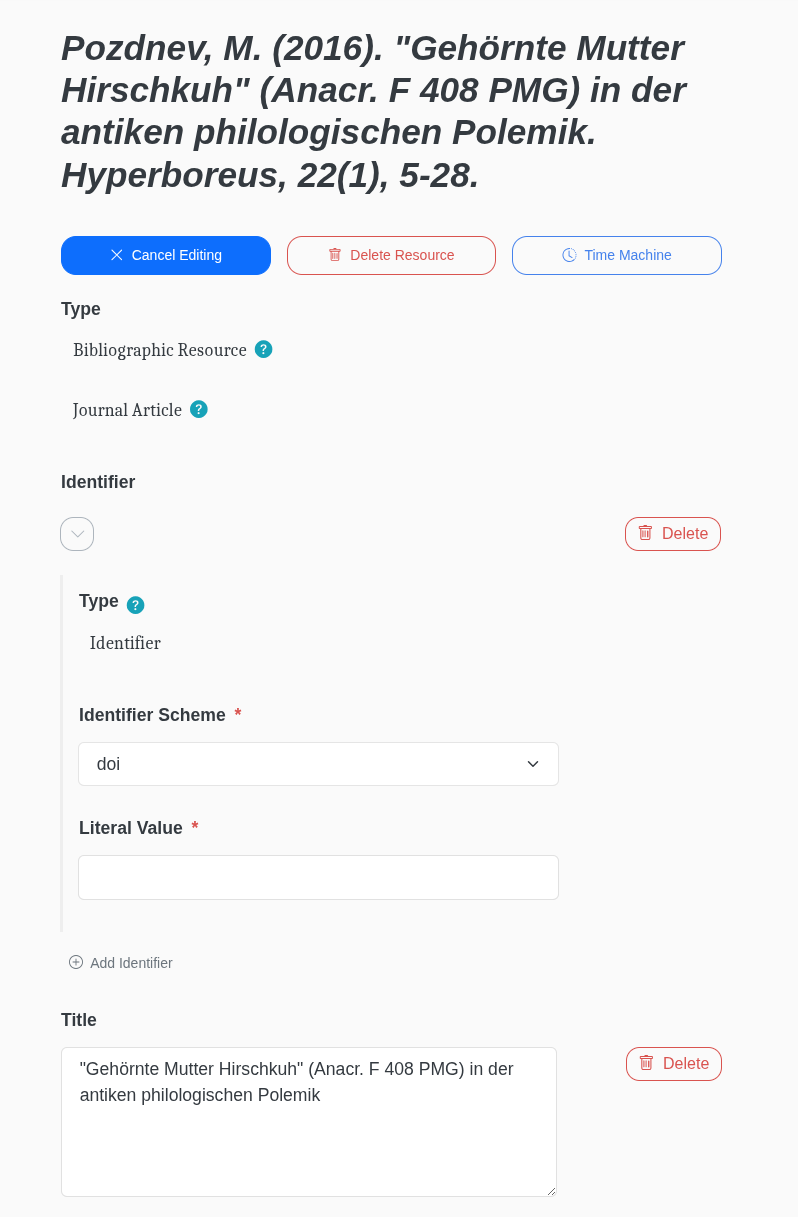}
\caption{Editing interface showing metadata fields and input controls}\label{fig:about}
\end{minipage}
\hfill
\begin{minipage}[t]{0.48\textwidth}
\centering
\includegraphics[width=\textwidth]{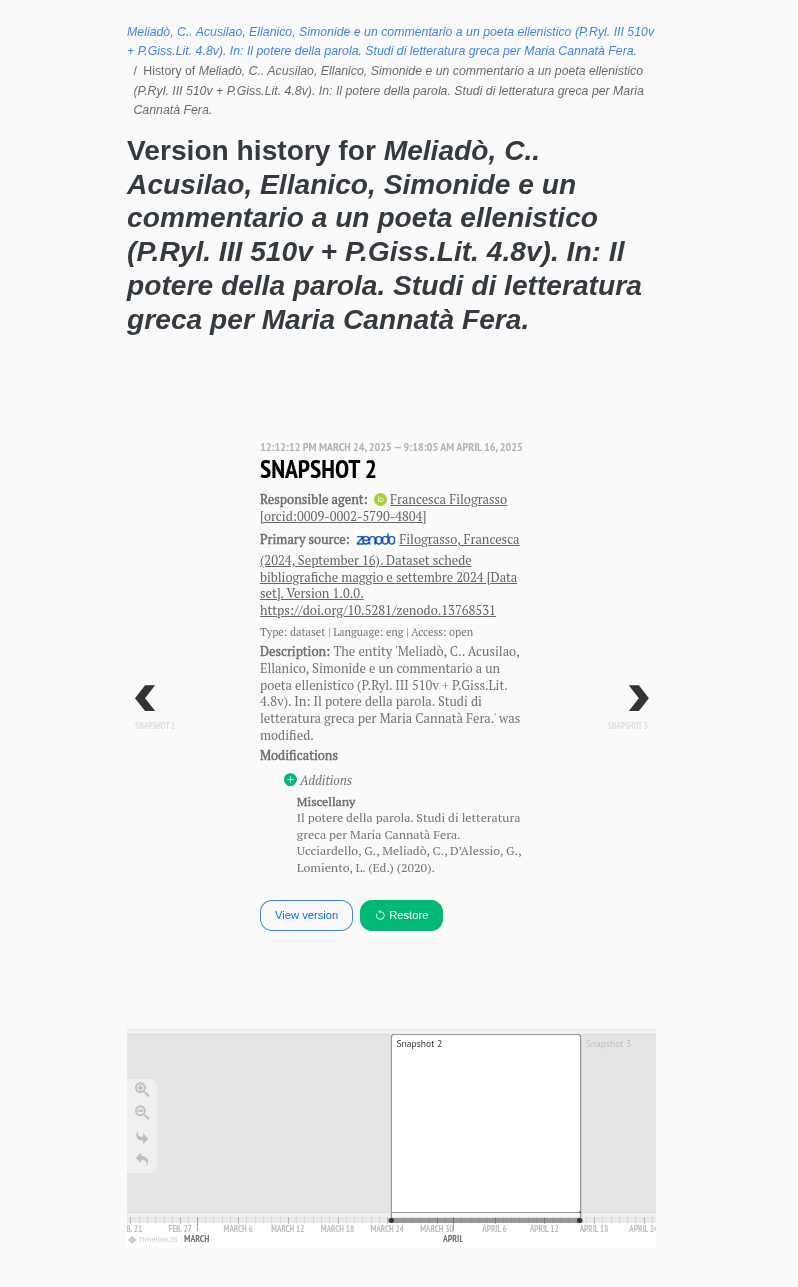}
\caption{Time Machine interface with version history and restoration controls}\label{fig:time_machine}
\end{minipage}
\end{figure}

The system automatically discovers entities from the connected triplestore based on their RDF class declarations. The editing view (Fig.~\ref{fig:about}) renders metadata fields through appropriate input controls: text fields, multiline text areas, date selectors, dropdown menus for controlled vocabularies, and tag inputs for keywords.

Form generation relies on two complementary configuration layers. SHACL shapes define the data model and drive widget selection: datatype declarations set which control appears (e.g. \texttt{xsd:date} produces a calendar, \texttt{xsd:gYear} a numeric field), cardinality constraints (\texttt{sh:minCount}, \texttt{sh:maxCount}) mark fields as required or repeatable, \texttt{sh:in} and \texttt{sh:or} constraints render as dropdown menus. Conditional constraints (\texttt{sh:condition}) enable context-sensitive validation~\cite{knublauchSHACLAdvancedFeatures2017}, e.g.\ enforcing DOI-specific patterns when the identifier scheme is DOI. Listing~\ref{lst:shacl} shows a shape that produces the identifier section in Fig.~\ref{fig:about}.

\begin{figure}[!ht]
\begin{lstlisting}[language=turtle, caption={SHACL shape for journal article identifiers. The \texttt{sh:in} constraint generates a dropdown menu, \texttt{sh:minCount} marks fields as required, and \texttt{sh:condition} activates validation rules only when a specific identifier scheme is selected (the example includes only DOI for brevity, but a production shape would define distinct patterns for each scheme).}, label={lst:shacl}]
schema:JournalArticleIdentifierShape
  a sh:NodeShape ;
  sh:targetClass datacite:Identifier ;
  sh:property [
    sh:path datacite:usesIdentifierScheme ;
    sh:minCount 1 ;
    sh:maxCount 1 ;
    sh:in ( datacite:doi ) ;
  ] ;
  sh:property [
    sh:path literal:hasLiteralValue ;
    sh:datatype xsd:string ;
    sh:minCount 1 ;
    sh:maxCount 1 ;
  ] ;
  sh:property [
    sh:path literal:hasLiteralValue ;
    sh:datatype xsd:string ;
    sh:minCount 1 ;
    sh:maxCount 1 ;
    sh:condition [
      sh:path datacite:usesIdentifierScheme ;
      sh:hasValue datacite:doi ;
    ] ;
    sh:pattern "^10\\.\\d{4,9}/[-._;()/:A-Z0-9]+$" ;
  ] .
\end{lstlisting}
\end{figure}

YAML files~\cite{ben-kikiYamlAintMarkup2005} govern presentation aspects outside SHACL's scope. SPARQL queries compute composite labels from the graph structure, for instance, assembling ``Volume 3 of Nature'' from a volume number and a journal title stored as separate triples. The YAML layer also controls which properties are visible and in what order, and allows per-field widget type overrides that replace the datatype-derived default (e.g.\ text area, date picker). Autocomplete can be configured for each form field, including the minimum character threshold before suggestions appear and whether the search targets entities of the same type or their parent resources, so that, for instance, users searching by an identifier value retrieve the bibliographic resource rather than the identifier entity itself. 

When the order of values matters, the system relies on intermediate proxy entities to maintain sequence information. For example, the author's ordering in OpenCitations data uses \texttt{pro:RoleInTime} instances, as defined in the Publishing Roles Ontology (PRO)~\cite{peroniScholarlyPublishingLinked2012}, as proxies: a configurable predicate chains these proxy entities so that each one points to the next. The interface exposes this ordering through drag-and-drop handles that update the chain when the user rearranges items. 

The same YAML layer configures duplicate detection for each entity type by specifying which properties to compare through boolean logic. For example, a person entity can be configured to flag duplicates that share the same full name, or the same combination of given name and family name, or the same identifier.

Each YAML configuration entry is bound to data either by RDF class or by SHACL shape. Shape-level binding allows distinct presentation configurations for entities that share the same class but conform to different shapes: for example, two shapes targeting \texttt{fabio:JournalIssue}~\cite{peroniFaBiOCiTOOntologies2012}, one for regular journal issues and one for special issues that additionally expose a title field, can each define a separate set of visible properties.

The SHACL and YAML configuration layers are independent and entirely optional. With no configuration at all, the system falls back to free-form property entry with display names auto-generated from URIs. Adding SHACL alone introduces structured forms with validation constraints while still deriving labels from URIs. Adding YAML alone keeps free-form entry but provides human-readable labels and controls property visibility and ordering. When both layers are present, SHACL governs form structure and validation while YAML refines the presentation. When a SHACL shape is defined for an entity type, only properties declared in that shape are available for creation and editing, ensuring data integrity with respect to the data model.

The YAML layer also supports virtual properties: fields that do not correspond to direct predicates in the data model but, when populated, automatically create and configure intermediate entities behind the scenes. Their purpose is to translate complex data models into interaction patterns aligned with users' mental models. Ontologies sometimes reify relationships as first-class entities to attach additional attributes. For instance, OpenCitations adopts the Citation Typing Ontology (CiTO)~\cite{peroniFaBiOCiTOOntologies2012} to model citations as instances of a dedicated class linking a citing entity to a cited entity through separate properties. A user's mental model of this relationship is simply ``I want to cite this paper'', not ``I want to create a Citation entity that links my article to that paper''. A virtual property labeled ``cites'' lets users select the target resource from a form field as though the relationship were direct; the system then creates the intermediate entity (i.e. the citation), assigns it the correct type (the class \texttt{cito:Citation}), and links it to both the current resource and the selected target (via the object properties \texttt{cito:hasCitingEntity} and \texttt{cito:hasCitedEntity}, respectively), without exposing this structure to the user. Each virtual property definition targets a specific SHACL shape, from which the system derives how to create and link the intermediate entity.

When editing operations sever relationships, the system applies configurable strategies for the affected entities. Orphaned entities that are no longer referenced by any other resource and proxy entities whose parent relationship has been removed can be handled by prompting the user for a decision, automatically deleting the entity, or retaining it in the dataset, depending on the deployment configuration.

Finally, creating new entities requires minting new IRIs. A built-in generator produces IRIs by appending a universally unique identifier (UUID) to a base namespace out of the box, so a basic deployment needs no code for this. Institutions that need their own naming conventions can provide them through a pluggable Python class.

Adapting HERITRACE to a new domain is configuration-driven: SHACL shapes and YAML display rules govern the data model and presentation, with no changes to the application code; the only optional code is a Python class for custom IRI schemes. At a minimum, deploying HERITRACE on a new collection requires only connecting it to a SPARQL endpoint and launching the application through the provided Docker image.

\subsection{Provenance and Change Tracking}\label{subsec:provenance}

Provenance tracking is automatic and mandatory: every creation, modification, deletion, and merge generates provenance records without requiring any explicit user action. The system implements snapshot-based versioning using the method described in the OpenCitations Data Model (OCDM)~\cite{daquinoOpenCitationsDataModel2020,massariRepresentingProvenanceTrack2025}. Each entity state is captured as a \texttt{prov:Entity} snapshot connected to the entity via \texttt{prov:specializationOf}, recording temporal boundaries, attribution, primary source, and predecessor chain through standard PROV Ontology~\cite{leboPROVOPROVOntology2013} properties. In addition, each snapshot stores a SPARQL \texttt{INSERT}/\texttt{DELETE} delta via the OpenCitations Ontology data property \texttt{oco:hasUpdateQuery}~\cite{peroniDocumentinspiredWayTracking2016}, capturing the exact triples changed. Every operation is non-destructive: changes produce new snapshots rather than overwriting existing data or provenance records.

The Time Machine interface of HERITRACE (Fig.~\ref{fig:time_machine}) presents the modification history of each entity as a chronological sequence of snapshots, displaying timestamps, responsible agents, primary sources, and change descriptions. Restoration replays the recorded update queries, reconstructing the entity together with the resources that the same tracked operations modified, and is itself recorded as a new snapshot. Deleted entities receive a final snapshot with a \texttt{prov:invalidatedAtTime} value; the Time Vault interface surfaces these entities, identifiable by the presence of both a generation time and an invalidation time, and allows them to be restored.

Merging entities is also non-destructive: it generates a new snapshot for the surviving entity recording the incorporated properties, while the absorbed entity's final snapshot is invalidated with a \texttt{prov:invalidatedAtTime} value. Provenance records connect the absorbed entity's snapshot to the surviving entity through \texttt{prov:wasDerivedFrom}.

To materialize past versions of entities, HERITRACE reuses the Time Agnostic Library~\cite{massariTimeTravelKnowledge2026}, a Python library for performing temporal SPARQL queries live on any SPARQL-compliant triplestore. When HERITRACE is deployed on a pre-existing RDF collection, the configuration accepts baseline values: a default primary source and the creation timestamp of the imported data, establishing a starting point for subsequent change tracking.

\section{System Assessment and Adoption}\label{sec:evaluation}

\subsection{Functional Comparison}\label{subsec:functional}

Table~\ref{tab:system_comparison} summarizes the comparison of the systems reviewed in Section~\ref{sec:relatedworks} against the five requirements (R1--R5) introduced in Section~\ref{sec:architecture}. According to our analysis, no existing system supports RDF-based change tracking (R4). HERITRACE, to the best of our knowledge, is the first RDF editing tool to record modifications as versioned snapshots natively in RDF. Beyond this novel capability, HERITRACE is the only system that simultaneously satisfies all five requirements.

\begin{table}[!htbp]
\centering
\caption{Comparison of semantic data management systems against requirements R1--R5}\label{tab:system_comparison}
\footnotesize
\begin{tabularx}{\textwidth}{l *{5}{>{\centering\arraybackslash}X}}
\hline
\textbf{System} & \textbf{User-} & \textbf{Admin-} & \textbf{RDF prov.} & \textbf{RDF change} & \textbf{Direct} \\
 & \textbf{friendly} & \textbf{friendly} & \textbf{mgmt} & \textbf{track.} & \textbf{RDF imp.} \\
\hline
OmekaS & \checkmark & \checkmark & & & \\
Semantic MediaWiki & \checkmark & \checkmark & & & \\
ResearchSpace & \checkmark & & & & \checkmark \\
CLEF & \checkmark & \checkmark & \checkmark & & \\
Wikibase & \checkmark & \checkmark & & & \\
Sinopia & \checkmark & \checkmark & & & \checkmark \\
Schímatos & & \checkmark & & & \checkmark \\
\textbf{HERITRACE} & \checkmark & \checkmark & \checkmark & \checkmark & \checkmark \\
\hline
\end{tabularx}
\end{table}

\subsection{Usability Study}\label{subsec:userstudy}

Prior to the formal evaluation, guerrilla testing~\cite{nielsenUsabilityEngineering1993} was conducted during the first production deployment of HERITRACE to surface initial issues and requirements from a real use case. In particular, we have deployed HERITRACE in ParaText~\cite{filograssoHERITRACEActionParaText2025}, an Italian inter-institutional research project on ancient Greek exegetical traditions whose bibliographical database requires domain-specific classifications that generalist resources such as the Ann{\'e}e Philologique~\cite{claymanDigitizationAnneePhilologique2021} do not represent. Classical philologists without Semantic Web expertise curate this data through HERITRACE, providing a realistic setting for identifying usability issues during routine scholarly work. Informal evaluations with these scholars identified interface problems that were iteratively addressed before a new controlled study, described below.

Building on these preliminary findings, such a formal study assessed two usability dimensions: whether domain experts without Semantic Web expertise can effectively curate RDF data through form-based interfaces, and whether technical staff with varying levels of SHACL experience can configure the system for new domains. A quantitative and qualitative evaluation of HERITRACE was conducted with two user populations: 9 end users (domain experts in bibliographic metadata) and 10 technicians (technical staff with Semantic Web knowledge). Among end users, 88.9\% had no prior HERITRACE experience. Among technicians, 90\% had no prior HERITRACE experience and 40\% had no prior SHACL experience. Participants completed self-guided sessions of up to 60 minutes using Docker packages with pre-configured HERITRACE instances on a 100-article subset of OpenCitations Meta~\cite{massariOpenCitationsMeta2024}. This setting, where participants worked independently in their own environments, maximizes ecological validity~\cite{kiefferECOVALFrameworkIncreasing2015}.

End users completed four metadata management tasks: (1)~editing an existing publication (adding an author and reordering the author sequence), (2)~merging duplicate author entities, (3)~restoring a previous version through the HERITRACE Time Machine, and (4)~creating a new publication from scratch, including authors, identifiers, and hierarchical containment relationships (i.e. from article to journal). Technicians completed two configuration tasks: (1)~adding a new abstract property to the HERITRACE data model through SHACL shapes and (2)~configuring YAML display rules for that property.

Data collection employed screen and voice recordings, the System Usability Scale (SUS)~\cite{brookeSUSQuickDirty1996}, and written reflections. Quantitative analysis tracked task completion, duration, error severity, and SUS scores. Qualitative analysis implemented grounded theory~\cite{corbinBasicsQualitativeResearch2008}, a method that allows categories to emerge from the data rather than imposing predefined ones, building a theory grounded in participants' actual experiences. All evaluation materials have been published on Zenodo~\cite{massariHERITRACEUserTesting2026}.

\begin{figure}[!htbp]
\centering
\includegraphics[width=\textwidth]{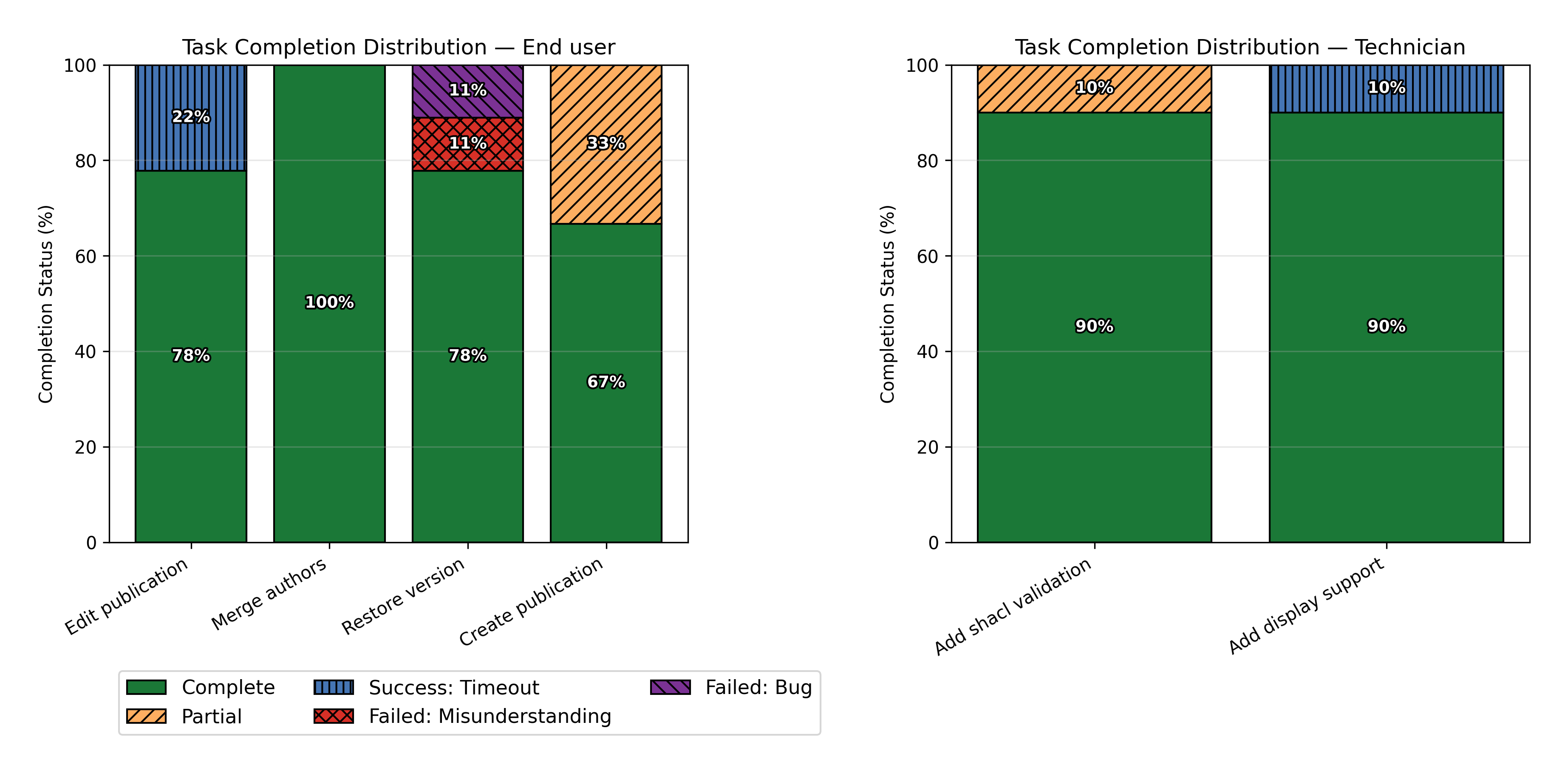}
\caption{Task completion status distribution by user type and task}\label{fig:success_rates}
\end{figure}

Figure~\ref{fig:success_rates} presents task completion distributions. End users achieved success rates of 78\% for editing, 100\% for merging, 78\% for version restoration, and 67\% for creating publications. The merge task demonstrated the highest success rate, with all 9 participants completing it. The two incomplete editing cases were timeouts rather than blocks: one participant browsed all records to find the publication due to the lack of a search feature; the other placed the new author in the wrong position, undid it through the Time Machine, and struggled with reordering. In version restoration, one participant misunderstood the feature and manually reverted the change instead of using the Time Machine, while the other was blocked by a bug that prevented the restore button from appearing. The create publication task yielded 6 complete and 3 partial outcomes. This task required building a containment hierarchy through nested collapsible sections (article, issue, volume, journal, and journal identifier), and the partial completions stemmed from participants losing track of previously entered data when outer sections collapsed. No participants abandoned any task due to insurmountable obstacles. Technicians achieved 90\% success for both SHACL validation and YAML display configuration. One partial completion occurred due to misreading the task requirements (setting the property as required instead of optional); one timeout occurred due to placing the configuration in the wrong YAML section.

\begin{figure}[!htbp]
\centering
\includegraphics[width=\textwidth]{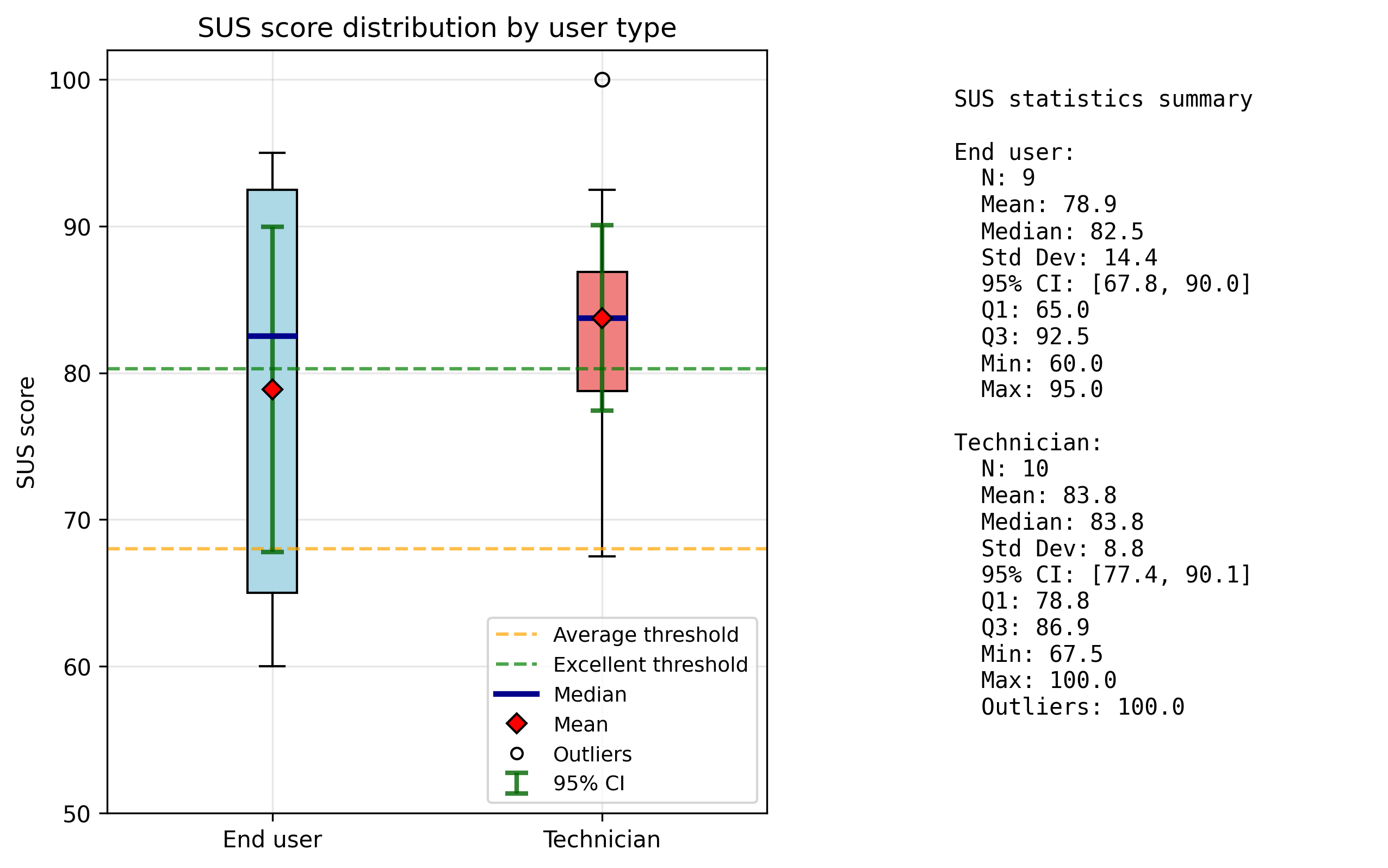}
\caption{SUS score distributions with benchmark thresholds for average (68) and excellent (80.3) usability~\cite{sauroQuantifyingUserExperience2016}}\label{fig:sus}
\end{figure}

Figure~\ref{fig:sus} presents SUS score distributions. SUS is treated as a unidimensional measure following~\cite{lewisRevisitingFactorStructure2017}. End users produced a mean SUS score of 78.9, placing usability in the above-average range~\cite{sauroQuantifyingUserExperience2016}. Technicians produced a mean of 83.8, exceeding the excellent threshold of 80.3.

Regarding the grounded theory analysis, among end users, the strongest positive responses concerned merge and duplicate management, the HERITRACE Time Machine as a safety net reducing anxiety about mistakes, and autocomplete for entity disambiguation. The most frequent negative categories were visual design problems with nested entity creation, where collapsible sections concealed previously entered data, and missing search functionality, reported by six of nine participants.

Among technicians, all participants employed pattern replication: locating existing SHACL or YAML configurations, copying them, and adapting them. This proved effective regardless of prior SHACL experience. Semantic reasoning and immediate visual feedback through automatic hot-reloading complemented this strategy. Substantive friction arose from the file-based configuration paradigm, and one technician questioned whether continuous technical support for configuration changes represents a sustainable model.

\subsection{Current and Potential Community Adoption}\label{subsec:adoption}

The ParaText deployment described in Section~\ref{subsec:userstudy} represents the first production adoption of HERITRACE. HERITRACE is also planned as the editing interface for OpenCitations. Separately, OpenCitations contributes to GRAPHIA~\cite{europeancommissionGRAPHIA2025}, a Horizon Europe project that aims to build a Knowledge Graph for the Social Sciences and Humanities (SSH), consolidating data scattered across multiple organizations. Within this consortium, OpenCitations is helping develop the SSH Citation Index, and HERITRACE will be tested as a possible tool for its curation layer.

Since the system operates on any SPARQL-accessible store without requiring data migration, its adoption potential extends beyond cultural heritage and scholarly metadata to any domain that maintains RDF data. More broadly, by abstracting RDF complexity behind form-based interfaces, HERITRACE lowers the barrier to adopting Semantic Web technologies.

\section{Discussion and Conclusion}\label{sec:limitations}

The usability study provided supporting evidence that both domain experts and technicians can work effectively with HERITRACE, while also exposing friction points that warrant refinement.

The end users in the controlled study were bibliographic metadata experts. This follows from the demonstration dataset, not from the tool: HERITRACE has no bibliography-specific logic, since every field a curator sees is generated from the SHACL shapes and YAML display rules. The evaluated tasks therefore exercise a domain-independent interaction model. However, a design argument is not evidence of usability. Thus, confirming that the model serves domain experts in other fields requires cross-domain evaluation that is left to future work.

Search functionality was deliberately excluded from HERITRACE, since searching RDF datasets, like visualizing and browsing resources, presents independent complexity. In OpenCitations, dedicated tools already address each problem separately: OSCAR~\cite{heibiEnablingTextSearch2019} provides configurable text-based search over SPARQL endpoints, while LUCINDA~\cite{ivanheibiOpencitationsLucinda2025} offers customizable resource visualization and browsing. HERITRACE concentrates on the editing problem. Currently, the three systems operate independently; connecting them to enable direct transitions from search results or resource browsing to the editing interface is a priority for future work.

Once users reach the editing interface, nested entity workflows introduce further friction. HERITRACE allows creating nested entities inline: when editing a publication, users can create the containing issue, volume, journal, and journal identifier within the same form, i.e. four nesting levels rendered as collapsible sections. The evaluation revealed two problems with this approach. First, collapsing a previously completed nested entity hides its content, forcing users to reopen sections to verify their input. Second, displaying all nesting levels within a single view produces cognitive overload. Two directions are under consideration: dialog-based sub-entity creation, isolating each nested entity in a separate modal window or tab so users only see the relevant context, and linearization, flattening the visual hierarchy so the containment structure remains transparent without nested collapsible sections.

On the other hand, the file-based configuration paradigm proved effective for technicians with Semantic Web expertise but raised sustainability concerns. Domain experts and technicians operate as distinct roles, so when the data model evolves (e.g.\ adding a class or modifying a property) the experts depend on technicians to update the SHACL and YAML configuration, a back-and-forth that can undermine the system's usability gains. This separation is not unique to HERITRACE: OmekaS, Semantic MediaWiki, ResearchSpace, and CLEF all require technical personnel at least for deployment, so eliminating the role entirely is unrealistic. A graphical configuration interface abstracting the underlying SHACL would help by exposing the supported constraints, validating fields immediately, and preventing the syntax-error crashes, while direct SHACL editing would remain for advanced users who need the full vocabulary.

Apart from the evaluation findings, the system's architectural scope also presents constraints. HERITRACE currently handles one triplestore at a time, but infrastructures such as OpenCitations manage interdependent collections served by different triplestores, where modifying an entity in one collection can affect references to it in another. Supporting such scenarios would require declaring dependency rules between collections, so that a modification in one triplestore automatically triggers the corresponding updates in the dependent ones, coordinated through atomic transactions to guarantee consistency.

The system also adopts a single-user editing model: Redis-based edit locks prevent concurrent modifications to the same entity rather than resolving conflicts. Real-time collaborative editing would require mechanisms for resolving concurrent RDF modifications. The provenance model already provides per-change attribution and version history, offering a foundation for future extension, but implementing such mechanisms exceeds the current scope.

The ordering mechanism described in Section~\ref{subsec:editing} currently requires intermediate proxy entities chained through a designated predicate. This covers ontologies like the Publishing Roles Ontology (PRO) but not alternative ordering strategies such as \texttt{rdf:List} or explicit numeric rank properties. Supporting multiple ordering paradigms would broaden the range of ontologies it can accommodate without custom adaptation.

A related constraint is that the system operates exclusively on RDF data, yet relational databases remain the most widely adopted data management systems~\cite{dbenginesRankingDatabaseManagement2026}, and many cultural heritage institutions maintain their collections in relational formats~\cite{stefanovaScalableReconstructionRDFarchived2013}. Keeping both representations synchronized requires bidirectional transformation. The RDB to RDF Mapping Language (R2RML)~\cite{w3cR2RMLRDBRDF2012} standardizes the forward direction, but inverting these mappings to propagate changes from a semantic editor back to the relational source remains an open problem. Preliminary work on R2RML mapping inversion~\cite{arcangelomassariArcangelo7KnowledgegraphsinversionV1202026} demonstrates that reconstruction is feasible when the mapping retains enough information to recover the original column values.

Beyond the data layer, user authentication also constrains deployability. HERITRACE authenticates users via ORCID OAuth~2.0, which provides both login and a persistent identifier for provenance attribution. This choice is appropriate for the academic contexts of both case studies, but ORCID adoption remains driven by journal submission requirements, with rates ranging from 17\% in visual and performing arts to 93\% in biological sciences among U.S.\ academic researchers~\cite{porterUnderstandingORCIDAdoption2025}. Curators, archivists, and librarians who are not also academic researchers are unlikely to hold ORCID identifiers. Alternative identifier registries relevant to the GLAM sector, such as ISNI~\cite{isoISO277292024} and VIAF~\cite{oclcVIAFVirtualInternational2026}, assign persistent identifiers but do not function as identity providers, since there is no OAuth flow for either. Extending HERITRACE to non-academic GLAM contexts would therefore require decoupling authentication from identifier assignment: an institutional identity provider for login, with persistent identifiers such as ISNI or VIAF URIs associated through a separate configuration step.

A further constraint concerns the configuration layer itself, as its representation is tool-specific. HERITRACE expresses presentation rules in YAML. Expressing them in RDF, for instance through YAML-LD~\cite{kelloggYAMLLD2026}, would turn the display configuration into a queryable data model reusable by other tools, extending to the configuration layer the FAIR treatment already applied to the data and its provenance.

The limitations discussed above define the boundaries of the current system rather than of the approach itself. The evaluation results suggest that usability and FAIR compliance in RDF curation need not be mutually exclusive: form-based interfaces driven by declarative constraints can make RDF data editable by domain experts without Semantic Web expertise. Thus, in its present form, HERITRACE can already be adopted in contexts where an RDF triplestore has been set up for data management, reducing the cognitive effort for domain experts to add, modify, and remove RDF data while maintaining provenance records and change history.

\paragraph*{Resource Availability Statement:} HERITRACE is available on GitHub, archived on Zenodo~\cite{arcangelomassariOpencitationsHeritraceV3002026} and Software Heritage Archive~\cite{massariHERITRACESoftwareHeritage2026} under the ISC license, registered on Wikidata (Q139571902), with Docker images on Docker Hub and GitHub Container Registry. Installation and usage documentation is available at \url{https://opencitations.github.io/heritrace/}. Evaluation data and materials discussed in this paper are on Zenodo~\cite{massariHERITRACEUserTesting2026} (CC BY 4.0).

\begin{credits}
\subsubsection{\ackname}
The authors thank Francesca Filograsso\orcidID{0009-0002-5790-4804} and Camillo Neri\orcidID{0000-0002-5041-4143} for their feedback during the development of HERITRACE within the ParaText project, and Bianca Gualandi\orcidID{0000-0001-8202-8493} for her assistance in preparing the informed consent documentation for the usability study. This work was funded by the European Union's Horizon Europe programme, Grant Agreement No.\@ 101188018 (GRAPHIA).
\subsubsection{\discintname}
AM is a software developer at OpenCitations and contributed to the development of several components of its current software infrastructure. SP is the Director of OpenCitations.
\subsubsection{Declaration of Use of Generative AI.}
No generative AI systems were used to design the study, generate or analyse data, produce results, or draft the scientific content of this manuscript. AI-assisted tools were used only for spell-checking and minor grammar suggestions on the English text.
\end{credits}

\bibliographystyle{splncs04}
\bibliography{references}

\end{document}